\documentclass[conference]{IEEEtran}
\usepackage{amsmath,amssymb,amsfonts}
\usepackage[ruled,vlined,linesnumbered]{algorithm2e}
\usepackage{graphicx}
\usepackage{mathrsfs}
\usepackage[margin=0.7in]{geometry}

\begin{document}

\title{\emph{DeepSlicing}: Deep Reinforcement Learning Assisted Resource Allocation for Network Slicing}

\author{\IEEEauthorblockN{Qiang Liu\IEEEauthorrefmark{1},
Tao Han\IEEEauthorrefmark{1}, Ning Zhang\IEEEauthorrefmark{2} and
Ye Wang\IEEEauthorrefmark{3}}
\IEEEauthorblockA{\IEEEauthorrefmark{1} Department of Electrical and Computer Engineering, The University of North Carolina at Charlotte, NC, United States\\
\IEEEauthorrefmark{2} Computer Science Department, Texas A\&M University at Corpus Christi, TX, United States\\
\IEEEauthorrefmark{3} School of Electronics and Information Engineering, Harbin Institute of Technology (Shenzhen), Guangdong, China\\
Email: \IEEEauthorrefmark{1}\{qliu12, tao.han\}@uncc.edu,
\IEEEauthorrefmark{2}ning.zhang@tamucc.edu,
\IEEEauthorrefmark{3}wangye83@hit.edu.cn}}


\maketitle

\begin{abstract}
Network slicing enables multiple virtual networks run on the same physical infrastructure to support various use cases in 5G and beyond. These use cases, however, have very diverse network resource demands, e.g., communication and computation, and various performance metrics such as latency and throughput. To effectively allocate network resources to slices, we propose DeepSlicing that integrates the alternating direction method of multipliers (ADMM) and deep reinforcement learning (DRL). DeepSlicing decomposes the network slicing problem into a master problem and several slave problems. The master problem is solved based on convex optimization and the slave problem is handled by DRL method which learns the optimal resource allocation policy. The performance of the proposed algorithm is validated through network simulations.
\end{abstract}


\section{Introduction}
\label{sec:introduction}
The emerging use cases and heterogeneous services, e.g., Internet of things (IoT), augmented/virtual reality (AR/VR) and vehicle-to-everything (V2X), drive the development and research on the 5th-generation mobile networks (5G) and beyond ~\cite{agiwal2016next}.
Unlike the conventional services, these services have highly diverse performance requirements such as bandwidth, delay, and reliability, which pose great challenges to network design in terms of scalability, availability, and cost-efficiency~\cite{foukas2017network}.

Leveraging network function virtualization, network slicing enables multiple virtual networks, i.e., network slices, run on top of a common physical network infrastructure~\cite{ordonez2017network}.
Each network slice can be tailored to meet the diverse network requirements of a specific use case. 
In network slicing, slice tenants have different service level agreements (SLAs) with the mobile network operator, e.g., slice throughput and end-to-end latency, and have full control of the operation of their slices, e.g., resource management and user admission control~\cite{liu2019direct}. 
The objective of network slicing for the network operator is to efficiently utilize network resources to maximize the overall network utility such as throughput, latency, and revenue and meet the SLAs of slices, which boils down to a network utility maximization problem.

In the literature, the network utility maximization (NUM) has been extensively studied~\cite{halabian2019distributed,salvat2018overbooking}. In these works, NUM is usually formulated as an optimization problem with given mathematical models and solved by various optimization methods, e.g., gradient descent methods.
However, the mathematical expression of utility functions of users can be very complicated and difficult to be obtained in real network  circumstances.
On one hand, the utility functions of users are affected by multiple factors, e.g., channel condition, user traffic, and network workload.
It is hard to obtain the closed-form mathematical models especially in highly dynamic mobile networks.
On the other hand, slice tenants have their own customized slice operation strategies, e.g, user admission and scheduling~\cite{ordonez2017network}.
These control strategies, which can be time-varying, change the utility of network slices.
As a result, it is impractical to assume the closed-form expression of utility functions in optimizing the resource allocation in network slicing.


Exploiting deep learning and deep reinforcement learning (DRL) for resource management in mobile networks has gained increasing research attentions~\cite{mao2016resource,liu2019network,xu2018experience}.
%
%
These works formulate the network utility maximization problem as a reinforcement learning problem and apply DRL techniques such as Deep-Q Learning to solve the problem.
It is shown that DRL obtains considerable improvement on the system performance in terms of throughput, latency, and utility.
However, these solutions are centralized network resource management which does not allow individual network slices to manage their own resources. Moreover, these solutions are designed for solving unconstrained optimization problems. As a result, they cannot guarantee the SLAs of network slices in resource allocation.
Thus, these solutions are not appropriate to solve the network slicing problem.

In this paper, we decompose the network slicing problem into a master problem and several slave problems by using the alternating direction method of multipliers (ADMM).
The master problem is solved by using convex optimization.
The slave problems are handled by the corresponding network slices so that the isolation among slice tenants can be ensured.
Since there is no closed-form expression of the utility functions of users in slave problems, we exploit the Deep Deterministic Policy Gradient (DDPG), which is a state-of-the-art DRL technique, to learn the optimal policy and allocate the resource to users accordingly.

The contributions of this paper are summarized as follows:
\begin{itemize}
    \item We design a new resource allocation method named DeepSlicing that integrates the ADMM method and deep reinforcement learning to dynamically slice the network without requiring the closed-form expression of the utility function of users in network slices. 
    \item We engineer a new machine learning algorithm based on DDPG with augmented state space and reward shaping to enable the coordination of the DDPG agents in solving the constrained resource allocation problem.
    \item We validate the performance of the DeepSlicing algorithm through extensive network simulations. The results show that the DeepSlicing algorithm significantly outperforms the baseline method and closely approaches the optimal solution. 
\end{itemize}

The remainder of this paper is organized as follows. In Section II, the
system model and problem formulation are presented. In Section III, we propose DeepSlicing that integrates the alternating direction method of multipliers and deep reinforcement learning. In Section IV, simulation results are provided to evaluate the performance of the proposed solution.
Finally, we conclude this paper in Section V.

\section{System Model and Problem Formulation}
We consider a base station (BS) with multiple network slices in a radio access network.
Network slices request radio resources to serve their own users. 
The mobile network operator manages the resource allocation to network slices and maximizes the overall utilities of all slices.

Let $\mathcal{I}$ and $\mathcal{K}_i$ be the set of network slices and users of the $i$th network slice, respectively.
Denote $x_{i,k}^{(t)}$ as the wireless data rate of the $k$th user in $i$th network slice, and let $\mathcal{X}_i^{(t)} = \{ x_{i,k}^{(t)}|\forall k \in \mathcal{K}_i\}$ as the set of resource allocation to the $i$th network slice at the $t$th time slot.
$\mathcal{X}^{(t)} = \{ \mathcal{X}_i^{(t)}|\forall i \in \mathcal{I}\}$ is the set of resource allocations at the $t$th time slot.
Then, utility of the $i$th network slice at the $t$th time slot is defined as
\begin{equation}
    \mathbf{U}_i^{(t)} =\sum\limits_{k \in \mathcal{K}_i}{w_{i,k} \mathbf{U}_{i,k}(x_{i,k}^{(t)})},
\end{equation}
where $\mathbf{U}_{i,k}(\cdot)$ is a non-decreasing utility function of the $k$th user in the $i$th slice at the $t$th time slot.
$w_{i,k}$ is the weight of the $k$th user in the $i$th slice. Here, $\mathbf{U}_{i,k}(\cdot)$ has no closed-form expression, so does $\mathbf{U}_i^{(t)}$.
Denote $R^{tot}$ and $\mathbf{U}_{i,k}^{\min}$ as the network capacity in terms of the data rate and the minimum utility requirement of the $k$th user, respectively.

The objective is to maximize the sum-utility of network slices which can expressed as $ \lim\limits_{T \to \infty} \frac{1}{T} \cdot \sum\nolimits_{t = 0}^{T} {\sum\nolimits_{i \in \mathcal{I}} \mathbf{U}_{i}^{(t)}} $.
Therefore, the network slicing problem is an infinite time horizon stochastic programming problem.
A common way to tackle the stochastic programming problem is to transform it into a problem with finite $\mathcal{T}$ time period~\cite{kall1994stochastic,salvat2018overbooking}. Therefore, we formulate the network slicing problem as
\begin{equation}
	\label{eq:problem}
	\begin{array}{*{20}{l}}
		{\mathscr{P}_1:}&{\max \limits_{\{ x_{i,k}^{(t)}\}}}&{ \sum\limits_{t \in \mathcal{T}} {\sum\limits_{i \in \mathcal{I}} \mathbf{U}_{i}^{(t)}} }\\
		{s.t.}&{C_1:}&{\;\sum \limits_{t \in \mathcal{T}} {\mathbf{U}_{i,k}^{(t)}} \ge \mathbf{U}_{i,k}^{\min}, \forall i \in \mathcal{I}, k \in \mathcal{K}_i},\\
		{}&{C_2:}&{\; 0 \le  x_{i,k}^{(t)} \le R^{tot}, \forall i\in \mathcal{I}, k \in \mathcal{K}_i,t \in \mathcal{T}},\\
		{}&{C_3:}&{\;\sum\limits_{i \in \mathcal{I}}\sum\limits_{k \in \mathcal{K}_i} x_{i,k}^{(t)} \le R^{tot}, \forall t \in \mathcal{T}}.
	\end{array}
\end{equation}
Here, constraints $C_1$ ensure that the minimum requirements of users' utilities are meet; constraints $C_2$ constrict the resource allocation to each user should not surplus the total amount of resource; constraints $C_3$ restrict that the amount of resource allocated to all users should not exceed the total amount of resource.

\section{Resource Allocation with Deep Reinforcement Learning}
\label{sec:algorithm}

In this section, we develop a resource allocation algorithm that effectively solves problem $\mathscr{P}_1$.
This problem is difficult to solve for two reasons.
First, utility functions of users $\mathbf{U}_{i,k}(\cdot), \forall i\in\mathcal{I},k\in\mathcal{K}_i$ have no closed-form expressions.
As a result, model-based algorithms, e.g., convex optimization and nonlinear programming, can not be used to solve the problem.
Second, the resource allocation to users in the problem are coupled by various constraints.

\begin{figure}[!t]
\centerline{\includegraphics[width=3.2in]{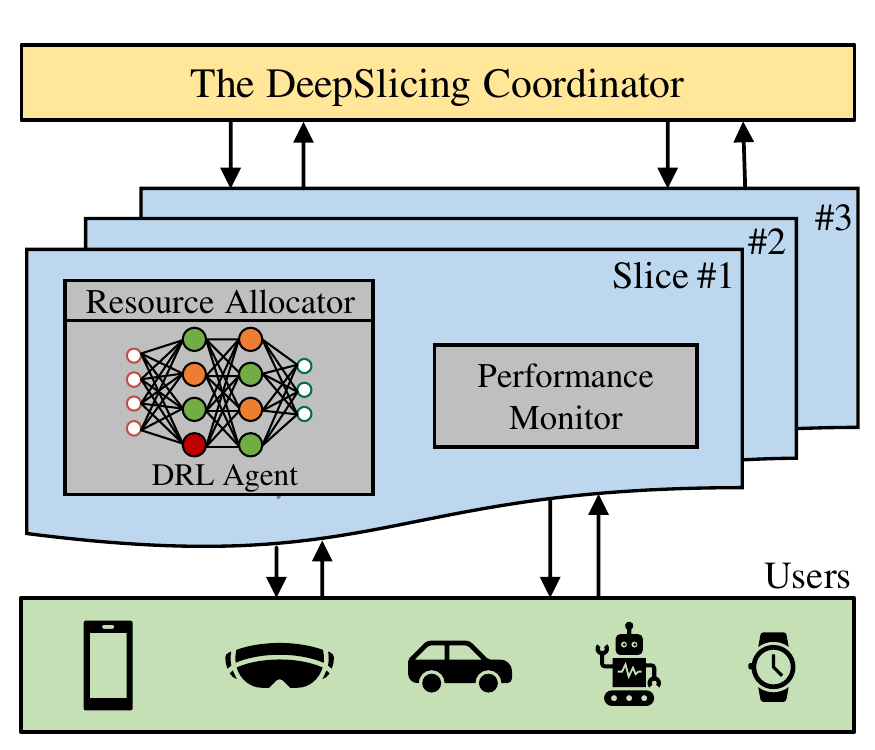}}
\caption{\small The overview of DeepSlicing.}
\label{fig:overview}
\end{figure}

To solve problem $\mathscr{P}_1$, we decompose it into a master problem and several slave problems by using the alternative direction method of multipliers (ADMM) method. 
As shown in Fig.~\ref{fig:overview}, the slave problems tackled in individual network slices focus on allocating the resources to users.
The master problem handled by the resource coordinator aims to coordinate the resource allocation among network slices by exchanging auxiliary and control variables with the slave problems.
As a result, problem $\mathscr{P}_1$ is resolved by iteratively solving the master problem and slave problems until the value of objective function converges.

The master problem is a standard quadratic programming problem and can be effectively solved by optimization tools, e.g., CVX~\cite{Convex2004Boyd}.
On solving the slave problem in each network slice, we leverage deep reinforcement learning (DRL) techniques to learn the optimal policy for the resource allocations to users.
In particular, we develop a Deep Deterministic Policy Gradient (DDPG)~\cite{lillicrap2015continuous} agent in every network slice to maximize its sum-utility while satisfying the minimum requirement of user utilities.

\subsection{Problem Decomposition}
To decompose the problem, we introduce an auxiliary variable $z_{i}^{(t)}$ and denote $\mathcal{Z}^{(t)} =\{ z_{i}^{(t)}|\forall i \in \mathcal{I}\}$. Then problem $\mathscr{P}_1$ is equivalent to
\begin{equation}
	\label{prb:transformed_problem}
	\begin{array}{*{20}{l}}
		{\mathscr{P}_2:}&{\max \limits_{\{ x_{i,k}^{(t)},z_{i}^{(t)}\}}}&{\sum\limits_{t \in \mathcal{T}} {\sum\limits_{i \in \mathcal{I}}\mathbf{U}_{i}^{(t)}} }\\
		{}&{s.t.}&{C_1,C_2,}\\
		{}&{C_3:}&{\; 0 \le  \sum\limits_{i \in \mathcal{I}} z_{i}^{(t)} \le R^{tot}}, \forall t \in \mathcal{T}\\
		{}&{C_4:}&{\sum\limits_{k \in \mathcal{K}_i} x_{i,k}^{(t)} = z_{i}^{(t)}, \forall i \in \mathcal{I}, t \in \mathcal{T}}.
	\end{array}
\end{equation}

In problem $\mathscr{P}_2$, there are two sets of variables, $\mathcal{X}^{(t)}$ and $\mathcal{Z}^{(t)}$, which are closely coupled by constraints $C_4$.
Based on the ADMM method~\cite{Convex2004Boyd}, we decompose problem $\mathscr{P}_2$ into a master problem that handles the update of variables $\mathcal{Z}^{(t)}$ and several slave problems that are responsible for optimizing variables $\mathcal{X}^{(t)}$.
Toward this end, we derive the augmented Lagrangian of problem $\mathscr{P}_2$ as
\begin{equation}
	\mathcal{L}_y = \sum\limits_{t \in {\mathcal{ T}}} {\sum\limits_{i \in {\mathcal{ I}}} {\left( {{\bf{U}}_i^{(t)} - \frac{\rho }{2}\left\| {\sum\limits_{k \in {{\mathcal{ K}}_i}} {x_{i,k}^{(t)}}  - z_i^{(t)} + y_i^{(t)}} \right\|_2^2} \right)} },
\end{equation}
where $\rho \geq 0$ is a positive constant, and $y_{i}^{(t)}$ is the scaled dual variable.
Here, the augmented Lagrangian incorporates the constraints $C_4$ that couple the variables $\mathcal{Z}^{(t)}$ and $\mathcal{X}^{(t)}$.
Then, problem $\mathscr{P}_2$ is solved by iteratively solving the following problems
\begin{align}
	\mathscr{P}_3: \;\;\;\;&{x_{i,k}^{(t+1)}}{= \arg \max_{x_{i,k}^{(t)} \in C_1, C_2} \mathcal{L}_y (x_{i,k}^{(t)},z_{i}^{(t)},y_{i}^{(t)}),}\label{x-update}\\
	\mathscr{P}_4: \;\;\;\;&{z_{i}^{(t+1)}}{= \arg \max_{z_{i}^{(t)} \in C_3} \mathcal{L}_y (x_{i,k}^{(t+1)},z_{i}^{(t)},y_{i}^{(t)}),} \label{z-update}
\end{align}
and updating the dual variables $\mathcal{Y}^{(t)} =\{ y_{i}^{(t)}|\forall i \in \mathcal{I}\}$ according to
\begin{equation}\label{u-update}
	{y_{i}^{(t+1)}} {= y_{i}^{(t)} +( \sum\limits_{k \in \mathcal{K}_i} x_{i,k}^{(t+1)}-z_{i}^{(t+1)})}.
\end{equation}
Here, $\mathscr{P}_3$ and $\mathscr{P}_4$ are the slave problems and master problem, respectively.
We first solve problem $\mathscr{P}_3$ with the auxiliary variables $\mathcal{Z}^{(t)}$ and dual variables $\mathcal{Y}^{(t)}$ derived from the last iteration.
We then solve problem $\mathscr{P}_4$ with the obtained variables $\mathcal{X}^{(t+1)}$ and the dual variables $\mathcal{Y}^{(t)}$ from solving problem $\mathscr{P}_3$.
We next update the dual variables $\mathcal{Y}^{(t)}$ by Eq.~\ref{u-update} with the obtained $\mathcal{X}^{(t+1)}$ and $\mathcal{Z}^{(t+1)}$ from solving problem $\mathscr{P}_3$ and problem $\mathscr{P}_4$, respectively. 

\subsection{Algorithm Design: Master Problem}
The master problem is responsible for optimizing auxiliary variables $\mathcal{Z}^{(t)}$ and updating dual variables $\mathcal{Y}^{(t)}$.
When we solve the master problem, variables $\mathcal{X}^{(t)}$ are known.
Therefore, problem $\mathscr{P}_4$ can be equivalently expressed as
\begin{equation}
	\begin{array}{*{20}{l}}
		{\mathscr{P}_5:}&{\min \limits_{\{ z_{i}^{(t)}\}}}&{ \sum\limits_{t \in {\mathcal{ T}}} \sum\limits_{i \in \mathcal{I}} \left\| {{\sum\limits_{k \in \mathcal{K}_i}x_{i,k}^{(t)}} - {z_{i}^{(t)}} + {y_{i}^{(t)}}} \right\|_2^2}\\
		{}&{s.t.}&{\; 0 \le  \sum\limits_{i \in \mathcal{I}} z_{i}^{(t)} \le R^{tot}, \forall t \in 
		\mathcal{T}}.
	\end{array}
\end{equation}
This is a standard quadratic programming problem which can be solved by using convex optimization tools, e.g., CVX~\cite{Convex2004Boyd}.
By solving problem $\mathscr{P}_5$, we obtain variables $\mathcal{Z}^{(t+1)}$, and then update dual variables $\mathcal{Y}^{(t)}$ according to Eq.~\ref{u-update}.


\subsection{Algorithm Design: Slave Problem}
Since the constraints in problem $\mathscr{P}_3$ (constraints $C_1$ and $C_2$) only restrict the resource allocation within a slice, problem $\mathscr{P}_3$ can be solved by each slice in parallel.
Therefore, problem $\mathscr{P}_3$ in each slice is written as
\begin{equation}
	\begin{array}{*{20}{l}}
		{\mathscr{P}_6:}&{\max \limits_{\{ x_{i,k}^{(t)}\}}}&{\sum\limits_{t \in \mathcal{T}} \left({{\mathbf{U}_{i}^{(t)}} - {  {\frac{\rho }{2}}  } \left\| {{\sum\limits_{k \in \mathcal{K}_i}x_{i,k}} - {z_{i}^{(t)}} + {y_{i}^{(t)}}} \right\|_2^2} \right)}\\
		{s.t.}&{C_1:}&{\; \sum \limits_{t \in \mathcal{T}} {\mathbf{U}_{i,k}^{(t)}} \ge \mathbf{U}_{i,k}^{\min}, \forall i \in \mathcal{I}, k \in \mathcal{K}_i},\\
		{}&{C_2:}&{\;0 \le  x_{i,k}^{(t)} \le R^{tot}, \forall  k \in \mathcal{K}_i,t \in 
		\mathcal{T}},
	\end{array}
	\label{prob:slave_problem}
\end{equation}
where $z_{i}^{(t)}$ and $y_{i}^{(t)}$ are derived from the solutions of problem $\mathscr{P}_5$.
The key challenge of solving the above problem is that the utility functions of users ${\mathbf{U}_{i}^{(t)}(\cdot)}, \forall i \in \mathcal{I}$ are without closed-form expressions.
To address this challenge, we design a new resource allocation algorithm based on deep reinforcement learning techniques.

\subsubsection{Deep Reinforcement Learning (DRL)}
We consider a general reinforcement learning setting where an agent interacts with an environment in discrete decision epochs.
At each decision epoch $t$, the agent observes a state $\mathbf{s}_{t}$, takes an action $\mathbf{a}_t$ based on its policy $\pi(\mathbf{s})$, and receives a reward $r(\mathbf{s}_t, \mathbf{a}_t)$. Then, the environment transits to the next state $\mathbf{s}_{t+1}$ based on the action taken by the agent.
The objective is to find the optimal policy $\pi^*(\mathbf{s})$ to map states to actions, that maximizes the discounted cumulative reward $R_0 =\sum\nolimits_{t=0}^{T} \gamma^t r(\mathbf{s}_t, \mathbf{a}_t)$, where $\gamma \in [0, 1)$ is the discounted factor. 

The challenges of applying DRL to solve problem $\mathscr{P}_6$ is two-fold.
First, it is challenging to user DRL to solve a constrained problem, especially the constraints are without closed-form expressions.
Second, it is difficult to integrate the DRL model into the master-slave architecture with considering varying exchanging variables effectively.

\subsubsection{DRL Design for Solving Slave Problem}

To address these challenges, we develop a new method to design state space and reward function specifically for solving problem $\mathscr{P}_6$.
First, we re-weight the constraints $C_1$ and incorporate it into the reward function of DRL so that the reward is affected by whether the constraints are met or not.
Second, we include $ {y_{i}^{(t)}} - {z_{i}^{(t)}} $ into the state space of DRL so that the agent can react under different auxiliary and dual variables.
The state space, action space and reward function are defined as follow.


\textbf{State Space}:
The state is composed of two parts: 1) the first part is $[\Delta^{t-1} + \mathbf{U}_{i,k}^{(t)} / \mathbf{U}_{i,k}^{\min}, \forall k \in \mathcal{K}_i ]$; 2) the second part is $[z_{i}^{(t)}-y_{i}^{(t)}]$. 
The first part represents how much utility the user obtained as compared to its minimum utility requirement.
The second part represents the auxiliary and dual variables from the master problem.
By augmenting the second part into the state space, the trained DRL agent is capable of allocating resource to users under different auxiliary and dual variables.
The state can be expressed as
\begin{equation}
\label{Eq:state}
    \mathbf{s}_t = \left[ \Delta^{t-1} + \mathbf{U}_{i,k}^{(t)} / \mathbf{U}_{i,k}^{\min}, \forall k \in \mathcal{K}_i; \;\;\; z_{i}^{(t)}-y_{i}^{(t)} \right],
\end{equation}
where $\Delta^{t-1} = \left({\sum \nolimits_{\tau=0}^{t-1} \mathbf{U}_{i,k}^{(\tau)}} \right) / \left({ \mathbf{U}_{i,k}^{\min}}\right)$.

\textbf{Action Space}:
The action is defined as the resource allocation to users in the network slice
\begin{equation}
\label{Eq:action}
    \mathbf{a}_t = [  x_{i,k}^{(t)}, \forall k \in \mathcal{K}_i  ].
\end{equation}

\textbf{Reward}:
We define the reward function as
\begin{align}
\label{Eq:reward}
    r(\mathbf{s}_t, \mathbf{a}_t) =& {\sum\limits_{k \in \mathcal{K}_i}{\left[\mathbf{U}_{i,k}^{(t)} +\beta \cdot \mathcal{H}\left({\mathbf{U}_{i,k}^{(t)}} - \mathbf{U}_{i,k}^{\min} / |\mathcal{T}|\right)\right]}} \\ \nonumber
    &- { {\frac{\rho }{2}}  } \left\| {{\sum\limits_{k \in \mathcal{K}_i}x_{i,k}^{(t)}} - {z_{i}^{(t)}} + {y_{i}^{(t)}}} \right\|_2^2,
\end{align}
where $\mathcal{H}(x) = (sigmoid(x) - 1)$ is a non-decreasing function, and $\beta$ is a positive constant. In particular, $\mathcal{H}(x) \to 0$ if $x \gg 0$, and $\mathcal{H}(x) \to -1$ if $x \ll 0$.
We design the reward function by integrating the objective function and constraints $C_1$ of of problem $\mathscr{P}_6$.
In this way, there will be a penalty added to the reward function if the minimum utility requirement of users are not satisfied.

Our objective is to develop a deep neural network that parameterized the policy of resource allocation to users.
Here, we use Deep Deterministic Policy Gradient (DDPG)~\cite{lillicrap2015continuous}, which is a state-of-the-art DRL technique, to train the deep neural network.
The DDPG is proposed by integrating the Deep Q-Network (DQN)~\cite{mnih2015human} and actor-critic method~\cite{konda2000actor} for solving problems with continuous and high-dimensional action spaces. 
In order to use DDPG, we design a 2-layer fully-connected neural network in both actor and critic networks, and there are 128 neurons in both layers with Leaky Recifier~\cite{goodfellow2016deep} activation functions.
In the output layer, we use $sigmoid$~\cite{goodfellow2016deep} as the activation functions to ensure that the resource allocation determined by the actions $\mathbf{a}_t$ will not exceed the total available resources.

\subsubsection{DRL Training Basis}
The basic idea of DDPG is to maintain a parameterized actor function $\pi(\mathbf{s}_t | \theta^\pi)$ and a parameterized critic function $Q(\mathbf{s}_t,\mathbf{a}_t | \theta^Q)$.
The critic function, which is implemented by using DQN, estimates the value function of state-action pairs.
The actor function specifies the current policy by mapping a state to a specific action.
It is implemented with another deep neural network which can be trained based on the Bellman equation~\cite{bellman1966dynamic}.

\textbf{DQN:}
The value function $ Q^\pi(\mathbf{s}_t, \mathbf{a}_t)$ is defined as the expected discounted cumulative reward if the agent starts with the state-action pair $(\mathbf{s}_t, \mathbf{a}_t)$ at decision epoch $t$ and then acts according to the policy $\pi$. The value function can be expressed as
\begin{equation}
    Q^\pi(\mathbf{s}_t, \mathbf{a}_t) = \mathop{\mathbb{E}}\limits_{\tau \sim \pi} {\left[ R_t | \mathbf{s}_t, \mathbf{a}_t \right]},
\end{equation}
where $R_t = \sum\nolimits_{k=t}^T \gamma^{(k-t)} r(\mathbf{s}_k, \mathbf{a}_k) $.
Based on the Bellman equation~\cite{bellman1966dynamic}, the optimal value function $Q^*(\mathbf{s}_t, \mathbf{a}_t)$ is
\begin{equation}
    Q^*(\mathbf{s}_t, \mathbf{a}_t) = r(\mathbf{s}_t, \mathbf{a}_t) + \gamma \max\limits_{\mathbf{a}_{t+1}}Q^*(\mathbf{s}_{t+1}, \mathbf{a}_{t+1}).
\end{equation}

To obtain the optimal policy, DQN is trained by minimizing the mean-squared Bellman error (MSBE) as follow
\begin{equation}
    L(\theta^Q) = \mathop{\mathbb{E}}\limits_{(\mathbf{s}, \mathbf{a}, r, \mathbf{s}') \in \mathcal{D}}{\left[ {\left( g_t - Q(\mathbf{s}_t, \mathbf{a}_t | \theta^Q)\right)}^2\right]},
\end{equation}
where $\theta^Q$ are weights of the Q-network and $\mathcal{D}$ is a replay buffer.
$g_t$ is the target value estimated by a target network
\begin{equation}
    g_t = r(\mathbf{s}_t, \mathbf{a}_t) + \gamma \max\nolimits_{\mathbf{a}_{t+1}}Q(\mathbf{s}_{t+1}, \pi(\mathbf{s}_{t+1}|\theta^{\pi'}) | \theta^{Q'}),
\end{equation}
where $\theta^{Q'}$ are weights of the target network.
The target network has the same architecture with the Q-network and its weights $\theta^{Q'}$ are slowly updated to track that of Q-network.




\textbf{Actor-Critic Method:}
The actor can be trained by applying the chain rule to the expected cumulative reward $J$ with respect to the actor parameters $\theta^\pi$
\begin{align}
    \nabla_{\theta^\pi} J &\approx \mathbb{E}[\nabla_{\theta^\pi} Q(\mathbf{s},\mathbf{a} | \theta^Q) |_{\mathbf{s}=\mathbf{s}_t,\mathbf{a}=\pi(\mathbf{s}_t)|\theta^\pi} ]\\ \nonumber
    &= \mathbb{E}[\nabla_{\mathbf{a}} Q(\mathbf{s},\mathbf{a} | \theta^Q) |_{\mathbf{s}=\mathbf{s}_t,\mathbf{a}=\pi(\mathbf{s}_t)}  \cdot \nabla_{\theta^\pi} \pi (\mathbf{s}|\theta^\pi)|_{\mathbf{s}=\mathbf{s}_t}].
\end{align}

The pseudo code of the proposed deep reinforcement learning resource allocation (DeepSlicing) algorithm for solving the network slice resource allocation problem is presented in Alg.~\ref{alg:proposed}.
At the beginning, we initialize auxiliary variables $\mathcal{Z}^{(t)}$ and dual variables $\mathcal{Y}^{(t)}$.
The DDPG agent in each slice is executed individually to obtain the resource allocations $\mathcal{X}^{(t+1)}$.
Then, with $\mathcal{X}^{(t+1)}$, auxiliary variables $\mathcal{Z}^{(t+1)}$ are obtained by solving problem $\mathscr{P}_5$, and dual variables $\mathcal{Y}^{(t+1)}$ are updated according to Eq.~\ref{u-update}.
	
	
	
	
	
	


\begin{algorithm}[!t]
	\caption{The DeepSlicing Algorithm}\label{alg:proposed}

	\KwIn{$\mathbf{U}^{\min}_{i,k}$, $\forall i \in \mathcal{I},k\in\mathcal{K}$, $R^{tot}$, $\rho$, $\eta$. }
	\KwOut{$x_{i,k}$, $\forall i\in\mathcal{I}, k\in\mathcal{K}$.}
	$t \gets 0$\;
    Initialize ${z_{i}^{(t)}}$ and ${y_{i}^{(t)}}$ randomly\; 
	\While{True}
	{
    	$/**\;optimize\;\mathcal{X}\; in\; each\; slave\; problem\; **/$\;
    	\For{$i \in \mathcal{I}$}
    	{
    	    $x_{i,k}^{(t+1)}$, $\forall k\in\mathcal{K}_i \gets$ the $i$th DPPG agents\;
    	}
    	
    	$/**\;update\;\mathcal{Z}\; in\; the\; master\; problem\; **/$\;
    	${z_{i}^{(t+1)}} \gets \arg \max\limits_{z_{i}^{(t)} \in C_1} \mathcal{L}_y (x_{i,k}^{(t+1)},z_{i}^{(t)},y_{i}^{(t)})$\;
    	$/**\;update\; dual \; variable\; \mathcal{Y} **/$\;
    	${y_{i}^{(t+1)} \gets y_{i}^{(t)} +(\sum\nolimits_{k \in \mathcal{K}_i} x_{i,k}^{(t+1)}-z_{i}^{(t+1)})}$\;
    	
    	$/**\;determine\; algorithm\; convergence\;**/$\;
    	\If{$\sum\limits_{i \in \mathcal{I}}\left\|\sum\limits_{k \in \mathcal{K}_i} x_{i,k}^{(t+1)} - z_i^{(t+1)} + y_i^{(t+1)}\right\| \le \eta$}
    	{   \Return{$x_{i,k}^{(t+1)}$, $\forall i \in \mathcal{I},k\in\mathcal{K}_i$}\;
    	}
    	$t \gets t + 1$\;
	}
\end{algorithm}

\section{Performance Evaluation}
In this section, we evaluate the performance of the DeepSlicing algorithm with network simulations.
In the simulation, we have 3 network slices, and each slice has 5 users.
To evaluate whether DeepSlicing can efficiently learn the utility function, We adopt the $\alpha$-fairness model, which is widely used in network utility maximization problems~\cite{xu2018experience,caballero2019network}, to calculate the utility of users in the simulations.
That is, $\mathbf{U}_{i,k}(x_{i,k})={x_{i,k}^{1-\alpha_{i,k}}}/\left({1-\alpha_{i,k}}\right)$, where $\alpha_{i,k} \in [0, 1]$ are randomly generated for users. Here, the model is only for calculating the utility and not seen by the DRL agent in the DeepSlicing algorithm for the resource allocation. 
The weights of users, $w_{i,k}$, are uniformly distributed between 0 and 1.
The minimum utility requirements of all users $\mathbf{U}_{i,k}^{\min}, \forall i \in \mathcal{I}, k \in \mathcal{K}_i$, are 2.
The total amount of resources $R^{tot}$ is 100, and $\rho=1.0$.

We implement a DDPG agent for solving the slave problem in each network slice using Tensorflow 1.10~\cite{abadi2016tensorflow}.
On training the DDPG agents, we conduct extensive and empirical tuning on the hyper-parameters.
The learning rates of both actor and critic networks are 0.001. The batch size is 1000.
The discounted factor for cumulative reward is $\gamma=0.99$.
We add the decaying Gaussian noise on actions $\mathbf{a}_t$ during the training phase for balancing the exploitation and exploration.
The noise starts from $\mathcal{N}(0,R^{tot})$ and decays with factor 0.9999 per update step.
The weight $\beta$ for the function $\mathcal{H}(\cdot)$ is 20.
During the training phase of a DDPG agent, we randomly generate $z_{i}^{(t)}-y_{i}^{(t)} $ between 0 and $R^{tot}$ to train the agent under different auxiliary and dual variables from the master problem.


We compare the DeepSlicing algorithm with the following algorithms:
\begin{itemize}
    \item ADMM with an optimization solver (\textbf{ADMMS}): We propose the ADMMS algorithm follows the procedures of DeepSlicing algorithm on solving problem $\mathscr{P}_1$ but replaces the DDPG agents in each slice with an optimization solver $fmincon$ in Matlab~\cite{MatlabR2019a}. The ADMMS algorithm is impractical in a real system because it requires the accurate model of utility functions of users.
    
    \item Static Resource Allocation (\textbf{SRA}): The SRA algorithm allocates the total resources to all network slices evenly, and slices equally share their resources to its users.
\end{itemize}

\begin{figure}[!t]
\centerline{\includegraphics[width=3.2in]{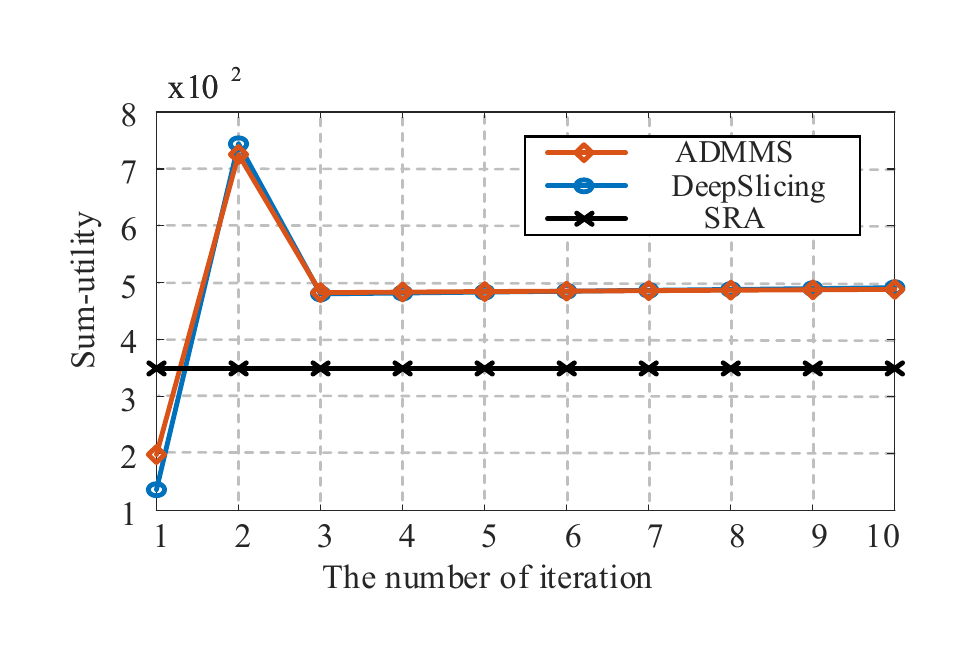}}
\vspace{-0.25in}
\caption{The convergence performance of the algorithms.}
\label{fig:combination1}
\vspace{-0.20in}
\end{figure}

\begin{figure}[!t]
\centerline{\includegraphics[width=3.2in]{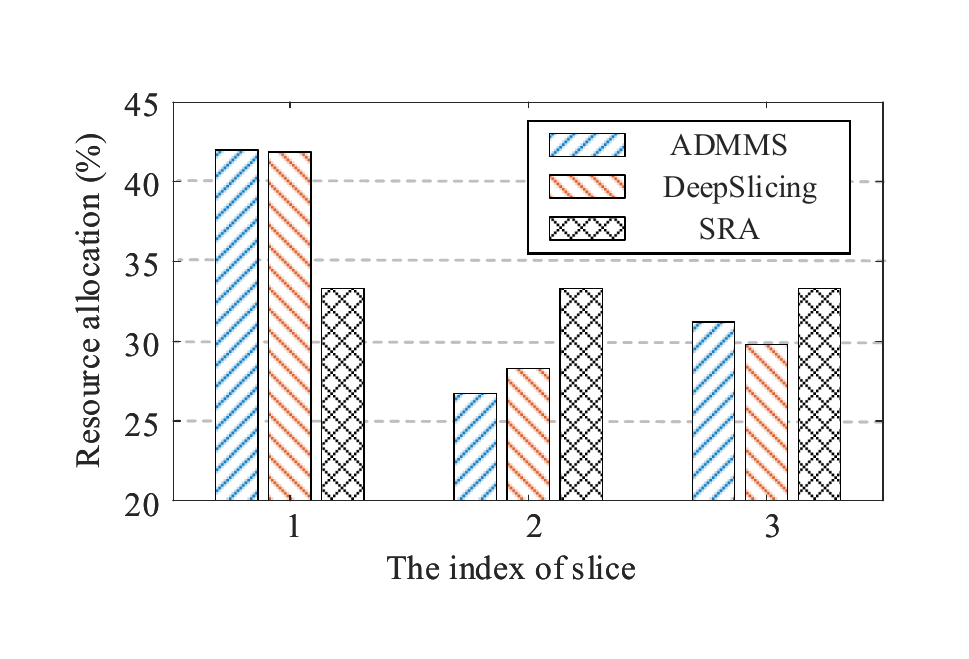}}
\vspace{-0.25in}
\caption{The resource allocation of the algorithms.}
\label{fig:combination2}
\vspace{-0.20in}
\end{figure}

\begin{figure}[!t]
\vspace{-0.1in}
\centerline{\includegraphics[width=3.2in]{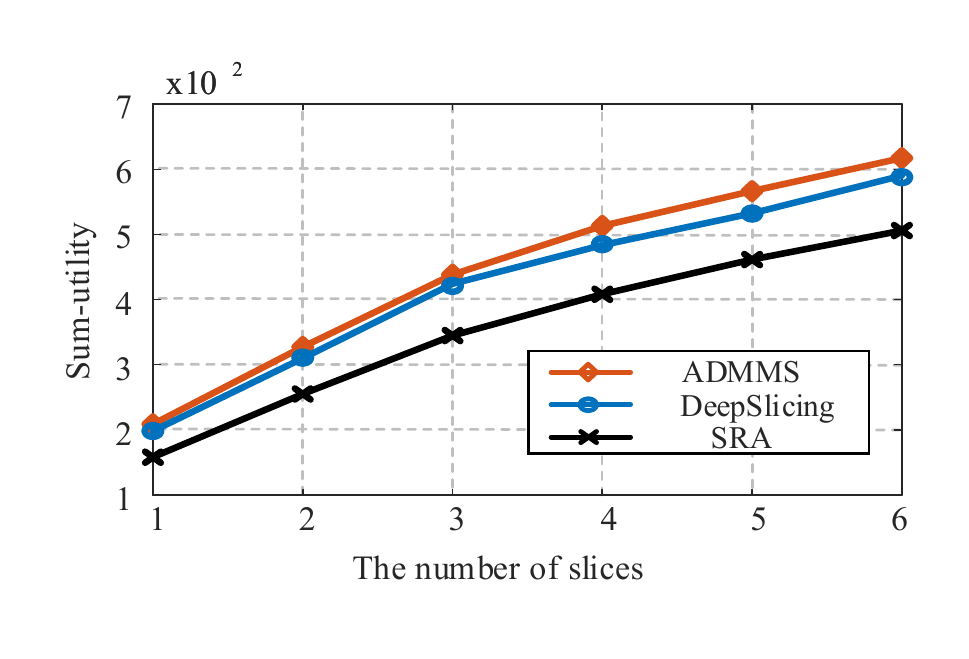}}
\vspace{-0.25in}
\caption{The sum-utility of the algorithms vs. the number of slices.}
\label{fig:combination3}
\vspace{-0.20in}
\end{figure}


\textbf{Convergence:}
Fig.~\ref{fig:combination1} shows the sum-utility versus the number of iterations.
Both the DeepSlicing and ADMMS algorithm converge in several iterations and have nearly the same sum-utility after the convergence.
During the iterations, the resource coordinator exchanges varying auxiliary variables and dual variables $z_i^{(t)} - y_i^{(t)}, \forall i \in \mathcal{I}$ with the DDPG agents in each network slice.
In every iteration, the DeepSlicing algorithm obtains almost the same performance as compared to the ADMMS algorithm.
This result proves that the DDPG agents are able to optimize the resource allocation for users under different auxiliary variables and dual variables.
The DeepSlicing algorithm obtains 1.42x sum-utility as compared to the SRA algorithm.

Fig.~\ref{fig:combination2} show the resource allocations of slices under different algorithms.
Instead of evenly allocating the resources to network slices and users, the DeepSlicing algorithm maximizes the sum-utility by learning to allocate resource to users in slice and adjusting resource allocation among slices.
For example, the DeepSlicing algorithm learns that slice 1 has higher utility per resource than other slices and hence allocates its major resources to slice 1. 
As we see that the DeepSlicing algorithm has a very similar resource allocation with the ADMMS algorithm (impractical in a real system) which solves the problem with optimization solvers.
This result validates the effectiveness of the DeepSlicing algorithm on utility maximization without closed-form utility functions.

\begin{figure}[!t]
\centerline{\includegraphics[width=3.2in]{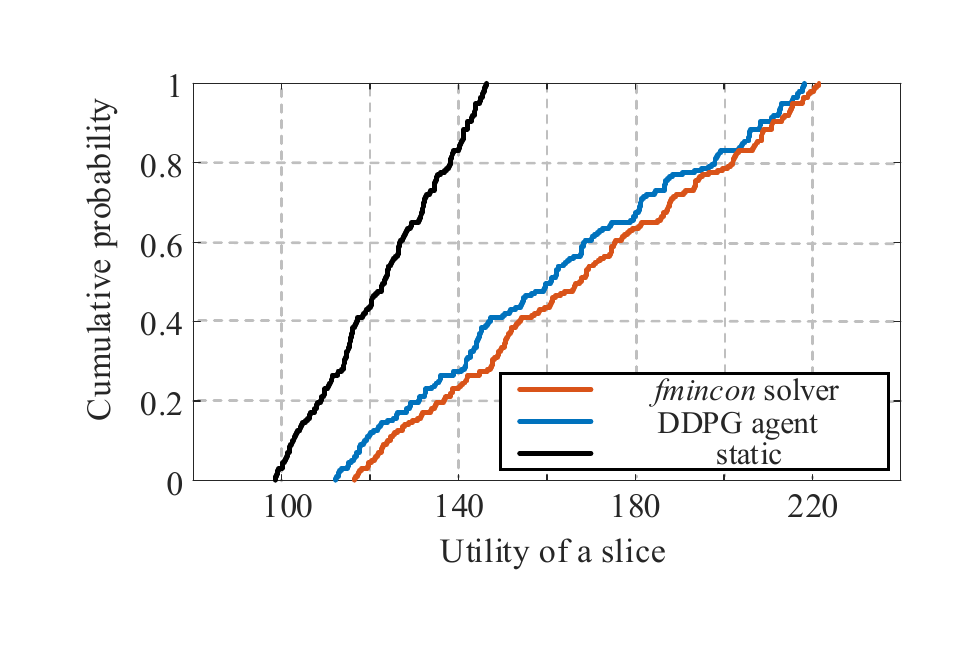}}
\vspace{-0.35in}
\caption{The cumulative probability of slice utility under the algorithms.}
\label{fig:combination4}
\vspace{-0.20in}
\end{figure}

\begin{figure}[!t]
\vspace{-0.1in}
\centerline{\includegraphics[width=3.2in]{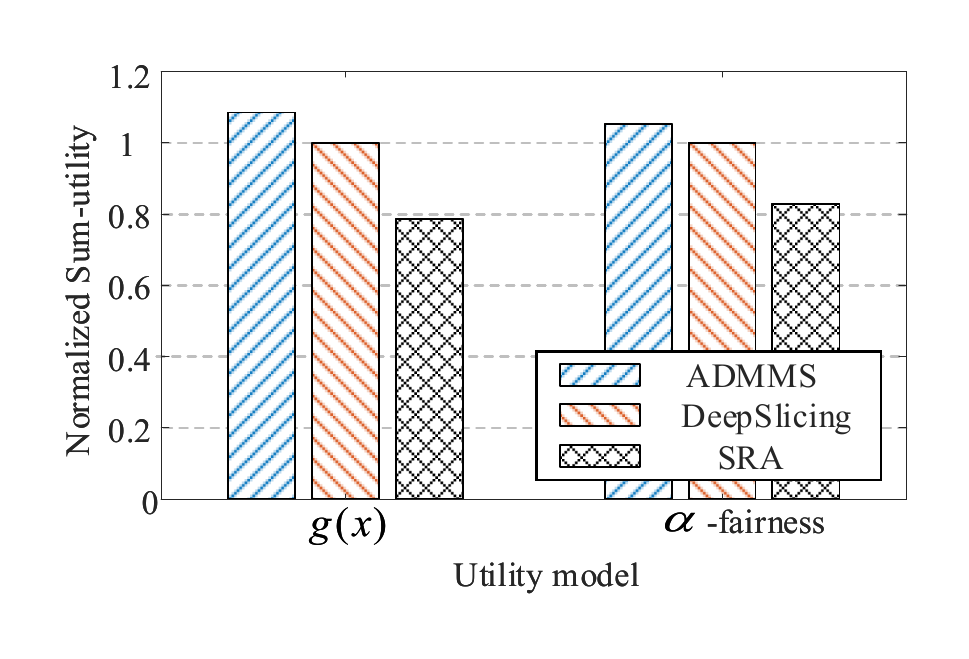}}
\vspace{-0.3in}
\caption{The normalized sum-utility vs. different utility functions.}
\label{fig:combination5}
\vspace{-0.20in}
\end{figure}

\textbf{Scalability:}
Fig.~\ref{fig:combination3} shows the sum-utility versus the number of network slices.
With the increment of number of network slices, the sum-utility increases accordingly.
The DeepSlicing algorithm obtains very similar performance of sum-utility as compared to the ADMMS algorithm, which validates the scalability of the DeepSlicing algorithm.
As the number of slices increases, the performance difference between the DeepSlicing and ADMMS algorithm slightly enlarges.
Since both the ADMMS and DeepSlicing algorithm follow the same procedures to solve problem $\mathscr{P}_1$, the only reason lies in the difference between trained DDPG agents and optimization solvers $fmincon$. 

To further study the effectiveness of the DDPG agent, we show the cumulative probability function (CDF) of utility obtained by different algorithms in solving a slave problem in Fig.~\ref{fig:combination4}.
The slave problem in a slice needs the auxiliary and dual variables $z_{i}^{(t)}-y_{i}^{(t)} $ from the master problem.
Here, we randomly generate the variables to evaluate the utility performance of different algorithms.
We can see there are slight differences between the DDPG agent and solver in terms of utility, which proves the effectiveness of the DDPG agent on resource allocation.
Although there is a negligible performance difference between the trained DDPG agent and optimization solver, it may still affect the performance of the algorithms.
With more network slices in the system, these performance differences may accumulate and thus hinder the DeepSlicing algorithm from obtaining the optimal performance.
Fortunately, this performance difference could be narrowed by introducing several techniques such as Hindsight~\cite{andrychowicz2017hindsight} when training the DDPG agents.



\textbf{Ability to Learning Utility Models:}
Fig.~\ref{fig:combination5} shows the normalized sum-utility performance of the algorithms under different utility models.
$g(x)=R^{tot}{\left(R^{tot}e^{-\alpha x}+1\right)^{-1}}$, which is non-decreasing and non-convex, is implemented as the other utility model.
The DeepSlicing algorithm substantially outperforms the SRA algorithm and closely approaches the ADMMS algorithm for both two utility models.
This shows that the deep reinforcement learning technique used in this paper is able to learn and optimize resource allocation even if the utility models are non-convex. 

\section{Conclusion}
In this paper, we have designed a new network slicing method named DeepSlicing. Aided by deep reinforcement learning, DeepSlicing learns how many resources are required by users in each slice to meet their QoS requirement and then optimizes the resource allocations accordingly. The performance of DeepSlicing has been validated in network simulations. The simulation results have showed that the performance of DeepSlicing approximates that of an optimization method which requires the exact model of users' QoS under different resource allocations. 

\bibliographystyle{IEEEtran}
\bibliography{ref/reference}

\end{document}